\documentclass[pdflatex,sn-mathphys-num]{sn-jnl}

\usepackage{graphicx}%
\usepackage{multirow}%
\usepackage{amsmath,amssymb,amsfonts}%
\usepackage{amsthm}%
\usepackage{mathrsfs}%
\usepackage[title]{appendix}%
\usepackage{xcolor}%
\usepackage{textcomp}%
\usepackage{manyfoot}%
\usepackage{booktabs}%
\usepackage{algorithm}%
\usepackage{placeins}
\usepackage{algorithmicx}%
\usepackage{algpseudocode}%
\usepackage{listings}%

\usepackage{tabularx}
\usepackage{array}

\newcommand{\yit}{{y_{it}}}
\newcommand{\rit}{{r_{it}}}
\newcommand{\Xit}{{X_{it}}}
\newcommand{\error}{{\epsilon_{it}}}

\newcommand{\alphai}{{\alpha_{i}}}

\usepackage{geometry}
\theoremstyle{thmstyleone}%

\theoremstyle{thmstyletwo}%

\theoremstyle{thmstylethree}%

\begin{document}

\title[Testing for racial bias\\using inconsistent perceptions of race]{\hspace{3cm}Testing for racial bias\\using inconsistent perceptions of race}

\author[1]{\fnm{Nora} \sur{Gera}}

\author[1,2]{\fnm{Emma} \sur{Pierson}}

\affil[1]{\orgdiv{Dept. of Computer Science}, \orgname{Cornell University}}

\affil[2]{\orgdiv{Dept. of Population Health Sciences}, \orgname{Weill Cornell Medical College}}

\abstract{Tests for racial bias commonly assess whether two people of different races are treated differently. A fundamental challenge is that, because two people may differ in many ways, factors besides race might explain differences in treatment. Here, we propose a test for bias which circumvents the difficulty of comparing two people by instead assessing whether the \emph{same person} is treated differently when their race is perceived differently. We apply our method to test for bias in police traffic stops, finding that the same driver is likelier to be searched or arrested by police when they are perceived as Hispanic than when they are perceived as white. Our test is broadly applicable to other datasets where race, gender, or other identity data are perceived rather than self-reported, and the same person is observed multiple times.}

\maketitle

\section{Introduction}\label{sec1}

Racial bias contributes to inequities in high-stakes settings including hiring, criminal justice, healthcare, and housing \cite{pager2008,kline2022systemic,pierson2020,simoiu2017,edelman2017racial,williams2015racial}. But rigorously quantifying bias is challenging for a fundamental reason: if two people of different races are treated differently, it is difficult to know whether this is because of their race or because of other confounding factors, an example of the omitted variable bias problem \cite{pierson2020, neil2019, ridgeway2006, knowles2001, simoiu2017, wilms2021}. To circumvent this issue, a long literature on quantifying racial bias has employed a host of strategies, including randomized controlled experiments (audit studies), natural experiments, and controlling for relevant factors~\cite{bertrand2004, grogger2006,gelman2007analysis,ridgeway2009doubly,arnold2020measuring,kline2022systemic,broockman2020natural,milkman2012temporal}. The common goal of these approaches is to remove or adjust for confounds so that remaining differences in how two people of different races are treated can be attributed to bias. 

	Here we take a fundamentally different approach. Instead of comparing two people of different races, we study the \emph{same person over time}. At first glance, this would seemingly remove the variation needed to measure bias, since a person’s race remains fixed over time. But, in fact, research has shown that perceptions of a single person's race can be inconsistent and fluid over time \cite{harris2002, maclin2001}. According to constructivist theories of race, observers rely on context clues --- e.g., the clothing someone is wearing~\cite{freeman2011looking} --- to perceive a person's race. Perceived race can differ from a person's self-identified race, and can change depending on the observer and other aspects of the situation~\cite{saperstein2006double, herman2010, harris2002, rose2023,baron2024unwarranted,finlay2024implications,samalik2023discrepancies,roth2016multiple,sen2016race}. 
 
 Our core approach is to use this inconsistency in perceived race to test for bias by assessing whether \emph{the same person is treated differently when their race is perceived differently}. For example, if the same person receives worse treatment when they are perceived as Hispanic than when they are perceived as white, this suggests anti-Hispanic bias. We motivate our test by building on a constructivist model of discrimination~\cite{rose2023}, as we discuss in more detail in $\S$\ref{sec2}.
 
 Our work builds on past studies which compare perceived to self-identified race in two important regards: first, we develop a test to specifically assess \emph{bias} in human decision-making (in contrast to past work, which focuses on how using different measures of race affects studies of inequality). Second, our method relies only on perceived race data, rather than self-identified race data, which is often unavailable. Our work also builds on a long literature using individual fixed effects to more reliably identify causal effects \cite{deneve2012, anekwe2015, xu2017, quintana2021, firebaugh2013, white2013, veenstra2020}. 
 
	We illustrate our method by applying it to test for racial bias in police traffic stops, using data from three states — Arizona, Colorado, and Texas — which provide the data necessary to track the same driver across multiple stops. Pooling data across all three states, 17\% of drivers are stopped multiple times, and 9\% of these multiply-stopped drivers have their race inconsistently perceived across stops (for example, Hispanic in one stop, and white in another). 
    We focus our analysis on drivers perceived as both white and Hispanic, since these drivers constitute 75\% of drivers whose race is inconsistently perceived.\footnote{Following previous literature on police discrimination, we use ``race'' throughout to refer to race/ethnicity~\cite{goel2016precinct,pierson2020,pierson2018fast,simoiu2017}. Hispanic identity is often considered an ethnicity and not a race~\cite{pew_hispanic}.} We find that these drivers are more likely to be searched or arrested by police when they are perceived as Hispanic than when they are perceived as white, suggesting bias against Hispanic drivers. A key advantage of our method, relative to past benchmarking tests for bias, is that it does not require us to control for all confounds which might influence search behavior for legitimate reasons, which are myriad and potentially unobserved; rather, it requires us only to control for confounds which 1) influence search behavior; 2) influence perceived race; 3) vary for the same person over time; and 4) do not suggest bias (e.g., the driver's perceived skin color plausibly satisfies the first three criteria, but not the fourth). We hence show that our core finding is robust to controlling for the variables which satisfy these criteria: for example, location, time, and individual officer effects. Beyond bias in policing, our core approach can be applied to test for bias in the many other settings --- including healthcare, survey, criminal justice, and child welfare datasets~\cite{saperstein2016making,roth2016multiple,samalik2023discrepancies,mcalpine2007agreement,saperstein2006double,hasnain2010barriers,finlay2024implications,baron2024unwarranted} --- where race, gender, or other measures of identity are perceived rather than self-reported and the same person is observed multiple times.

\section{Results}\label{sec2}

Our test quantifies whether the same person is treated differently when their race is perceived differently. We first describe the motivation for this test and then provide a mathematical description of how we implement it. 

\subsection{Motivation}

We motivate our test with a model similar to the constructivist model of discrimination proposed in~\cite{rose2023}. For concreteness, we describe the model in the context of our specific empirical setting --- assessing whether the same person is more likely to be searched by police when they are perceived as Hispanic than when they are perceived as white --- but our approach naturally extends to other decision-making settings. We assume that when an officer stops a driver, they observe features $X$ --- the appearance and behavior of the driver, type of car, time and date of the stop, and so on --- and, on the basis of those features, perceive the driver's race, $r(X) \in [0, 1]$, on a binary continuum which ranges from white ($r(X) = 0$) to Hispanic ($r(X) = 1$).\footnote{For simplicity, we follow~\cite{rose2023} in modeling race as a binary continuum, but our approach is also applicable to more than two racial categories.} After perceiving the driver's race, as in past models of police searches~\cite{simoiu2017,pierson2018fast,pierson2020}, the officer infers the probability the driver is carrying contraband --- $p(X, r) \in [0, 1]$ --- and searches the driver if this probability exceeds a threshold $t(X, r) \in [0, 1]$. Both the inferred probability and threshold may vary both by $r$ and by $X$. 

If we find that the same person is more likely to be searched by police when they are perceived as Hispanic than when they are perceived as white, three possibilities might explain this result. 

\begin{enumerate} 
\item \textbf{Police infer that the same person is more likely to be carrying contraband when they perceive them as Hispanic}. In other words, $p(X, r = 1) > p(X, r = 0)$. This constitutes \emph{statistical racial discrimination}~\cite{arrow1973discrimination}; which is generally illegal~\cite{starr2023statistical}, including in the police search context that we study~\cite{simoiu2017}. 
\item \textbf{Police apply a lower threshold for searching the same person when they perceive them as Hispanic}. In other words, $t(X, r = 1) < t(X, r = 0)$. This constitutes \emph{tasted-based racial discrimination}~\cite{arrow1973discrimination}, which is also illegal both in general and in the police search context that we study~\cite{simoiu2017}.
\item \textbf{Some aspect of $X$ changes for the same person over time and affects both their perceived race and the officer's search behavior.} This possibility requires further investigation, because there are a limited number of possibilities for time-varying $X$ which might not suggest bias on the part of decision-makers. For example, if the same driver interacts with both Hispanic and white police officers, and Hispanic officers are both likelier to perceive someone as Hispanic and likelier to conduct searches, drivers would be more likely to be searched when perceived as Hispanic even in the absence of bias by any individual officer; the race of the decision-maker thus acts as the time-varying component of $X$. Note that a major benefit of our method, in contrast to past benchmarking tests for bias, is that it does not require us to control for all $X$ which might influence search behavior for legitimate reasons~\citep{simoiu2017}. Rather, our test requires us only to control for the much more limited set of $X$ which 1) influence search behavior; 2) influence perceived race; 3) vary for the same person over time; and 4) do not suggest bias (e.g., the driver's perceived skin color plausibly satisfies the first three criteria, but not the fourth). In our sensitivity analyses below, we explicitly check for such non-bias explanations.
\end{enumerate}

\subsection{Empirical implementation}

As above, we introduce notation in the context of our specific empirical setting, but our approach naturally extends to other decision-making settings and to more than two race groups. Let $i$ index people and $t$ index time. Let $\yit \in \{0, 1\}$ be a binary variable indicating whether the person was searched by police; $\rit \in \{0, 1\}$ be a binary variable denoting whether their race was perceived as Hispanic ($\rit = 1)$ or white ($\rit = 0$); and $\Xit$ denote covariates which may vary by person and over time. We estimate the following model: 
\begin{equation*}
   \yit = \alphai + \beta \Xit + \delta \rit + \error
\end{equation*}
The observed data are $\{\yit, \Xit, \rit\}$; the parameters to be estimated are the person fixed effects $\alphai$, the coefficients on controls $\beta$, and the main parameter of interest $\delta$, which quantifies how the same person's likelihood of being searched differs when the perception of their race differs. 

In our primary specification, we estimate coefficients using a linear fixed effects model and report standard errors clustered at the driver level. Because our empirical setting analyzes a binary outcome variable $\yit$ (whether the police conduct a search), this corresponds to a linear probability model. Such models are often used in fixed effects settings with binary variables~\cite{goldin2000,huntington2021effect} due to their simplicity and interpretability. However, to confirm our results are robust to misspecification caused by the binary outcome, we additionally fit several models specifically designed for binary outcomes (Figures \ref{fig:feglm_no_bias_correct_search_rate_fig} and \ref{fig:cond_logit_search_rate_fig}). Our results remain similar with, and our framework naturally accommodates, these alternate specifications. 

\begin{table*}[ht]
\centering
\caption{Descriptive statistics of the dataset. Each section of the table reports statistics for increasingly nested subsets of the data: all drivers, multiply-stopped drivers, multiply-stopped drivers with inconsistently perceived race, and inconsistently-perceived drivers recorded specifically as white and Hispanic.} 
\label{tab:descriptive_stats}
\small
\begin{tabular*}{1.02\textwidth}{@{}p{6.3cm}llll@{}}
  \toprule
 & \textbf{Arizona} & \textbf{Colorado} & \textbf{Texas} & \textbf{Overall} \\ 
  \midrule
\toprule\multicolumn{5}{l}{\textbf{Full dataset}}\\
\toprule
Drivers & 1,836,709 & 1,548,561 & 1,510,143 & 4,895,413 \\ 
   \midrule
Stops & 2,214,995 & 2,146,534 & 1,760,100 & 6,121,629 \\ 
  \midrule
\toprule
\multicolumn{5}{l}{\textbf{Multiply-stopped drivers}}\\
\toprule
 Drivers & 275,412 & 345,774 & 184,257 & 805,443 \\ 
  \midrule
\% of all drivers & 15.0\% & 22.3\% & 12.2\% & 16.5\%\\\midrule
 Stops & 648,797 & 927,533 & 430,843 & 2,007,173 \\ 
  \midrule
\% of all stops & 29.3\% & 43.2\% & 24.5\% & 32.8\%\\ \midrule\toprule
\multicolumn{5}{l}{\textbf{Multiply-stopped drivers with inconsistently perceived race}}\\
\toprule
 Drivers & 30,065 & 32,565 & 9,473 & 72,103 \\ 
  \midrule
\% of all multiply-stopped drivers & 10.9\% & 9.4\% & 5.1\% & 9.0\%\\\midrule 
Stops & 81,625 & 100,760 & 23,480 & 205,865 \\ 
  \midrule
\% of all multiply-stopped driver stops & 12.6\% & 10.9\% & 5.4\% & 10.3\%\\ \midrule\toprule
\multicolumn{5}{l}{\textbf{Drivers perceived as both white and Hispanic}}\\
\toprule
 Drivers & 19,285 & 27,423 & 7,462 & 54,170 \\ 
  \midrule
\% of all inconsistently-perceived drivers & 64.1\% & 84.2\% & 78.8\% & 75.1\%\\ \midrule
 Stops & 52,771 & 86,177 & 18,807 & 157,755 \\ 
  \midrule
\% of all inconsistently-perceived driver stops & 64.7\% & 85.5\% & 80.1\% & 76.6\%\\ \midrule
 Search rate when perceived as Hispanic & 4.2\% & 0.5\% & 1.8\% & 1.9\% \\ 
   \midrule
 Search rate when perceived as white & 3.2\% & 0.4\% & 1.5\% & 1.5\% \\ 
   \bottomrule
\end{tabular*}
\end{table*}

\subsection{Policing data}

We use data and data processing code from the Open Policing Project~\cite{pierson2020}, a study of police traffic stops across the United States. We analyze state patrol traffic stop data from the three states --- Arizona, Colorado, and Texas --- which provide the information required to track the same driver across multiple stops, as our approach requires: for example, in Colorado, we link a driver across multiple stops using name and date-of-birth (see SI for further details). 17\% of drivers are stopped multiple times, and 9\% of these multiply-stopped drivers have their race inconsistently perceived across stops (for example, Hispanic in one stop, and white in another). The large majority of these inconsistently-perceived drivers (75\%) are recorded as both Hispanic and white (as opposed to some other combination of race groups); we thus filter for these drivers, who comprise our final analysis sample of 54,170 drivers and 157,755 stops. We use the \emph{search rate} --- i.e., a driver's probability of being searched after being stopped by the police --- as our metric of negative treatment, following past work~\cite{pierson2020,pierson2018fast,simoiu2017,anwar2006alternative,knowles2001} (we confirm that results remain similar when using the arrest rate as an alternate metric). Importantly, in all three states, the search rate is higher for our sample of inconsistently-perceived drivers when they are perceived as Hispanic than when they are perceived as white (Arizona: 4.2\% vs 3.2\%; Colorado: 0.5\% vs 0.4\%; Texas: 1.8\% vs 1.5\%). Table \ref{tab:descriptive_stats} provides descriptive statistics for the sample, and the SI fully describes all data processing procedures.

\subsection{Estimates of bias}

\begin{figure}[b!]
    \centering
    \includegraphics[width=0.75\linewidth]{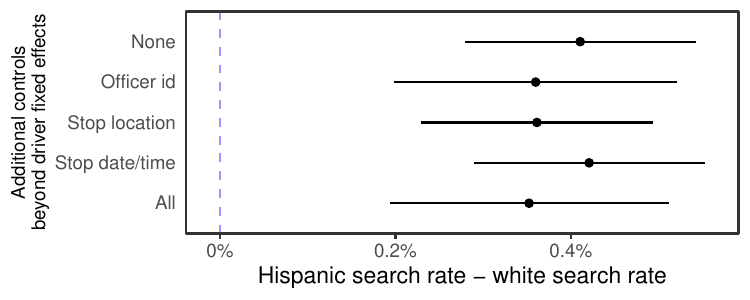}
    \caption{Estimated increase in the search rate when the same driver is perceived as Hispanic as opposed to white, using a linear probability model. All estimates include driver fixed effects. 95\% confidence intervals are plotted with standard errors clustered at the driver level. Estimates remain similar when including controls for officer identity, stop location, and stop date and time.}
    \label{fig:search_rate_fig}
\end{figure}

Figure \ref{fig:search_rate_fig} presents results from our primary specification, a linear probability model, pooling data across all three states and clustering standard errors at the driver level. With driver fixed effects, but no additional controls, we estimate that the same driver is 0.4 percentage points (pp) more likely to be searched when they are perceived as Hispanic than when they are white, suggesting bias against Hispanic drivers (95\% CI, 0.3 - 0.5 pp, $p < 10^{-4}$). This is not just a statistically significant but also a practically significant effect: it is 24\% of the average search rate across all states (1.7\%).

Results remain robust across three sets of sensitivity analyses. First, results remain robust to inclusion of additional controls. We assess the sensitivity of the results when including the battery of controls which are available across all three states and used in previous analyses of this dataset~\cite{pierson2017large}: stop location (encoded as county) and stop date and time (encoded as stop year, stop quarter, weekday, and stop hour binned into eight three-hour bins). Stop location might influence both racial perception (because driver demographics correlate with location)~\cite{finlay2024implications} and search behavior~\cite{goel2016precinct,simoiu2017,pierson2018fast,pierson2017large,pierson2020}; similarly, time of day might influence both racial perception (if, for example, lighting conditions differ~\cite{grogger2006,pierson2020}) and search behavior. Estimates remain similar (Figure \ref{fig:search_rate_fig}) when including these additional controls. We also assess whether our results remain robust when including a fixed effect for the officer conducting the stop, since officer identity might affect both search behavior and perception of the driver's racial identity; our estimates remain similar (Figure \ref{fig:search_rate_fig}). A final explanation for our results which does not necessarily suggest bias is that searches result in longer stops, giving the officer more time to perceive the race of the driver and altering their perception of the driver's race. We investigate this explanation using data from Arizona, the only state which records stop duration. The discrepancy in search rates remains large when we include controls for stop duration (0.6 pp, 95\% CI, 0.3 - 0.9 pp, $p < 10^{-4}$), suggesting this does not explain the results.  

Second, we examine the robustness of our results to an alternate definition of negative treatment --- in particular, whether the driver was arrested --- in the states for which arrest data is available, Arizona and Colorado. We find similar evidence of bias in arrest rates (Figure \ref{fig:az_co_arrest_rate_fig}): we estimate that the same driver is 0.4 pp (95\% CI, 0.2 - 0.6 pp, $p < 10^{-4}$) more likely to be arrested when they are perceived as Hispanic than when they are white. 

Third, we show that our results remain robust when using alternate statistical models designed for binary outcomes: in particular, a fixed effects generalized linear model with a logit link (Figure \ref{fig:feglm_no_bias_correct_search_rate_fig}) and a conditional logistic regression (Figure \ref{fig:cond_logit_search_rate_fig}). Collectively, these sensitivity analyses show that our finding of bias against Hispanic drivers remains robust across alternate sets of controls, outcome definitions, and statistical models.

\section{Discussion}\label{sec12}

We propose a test for bias which quantifies whether the same person is treated differently when their race is perceived differently. We apply our method to assess bias in searches after state patrol stops across three states, finding that the same driver is more likely to be searched or arrested when they are perceived as Hispanic than when they are perceived as white. 

Several points are important to keep in mind when interpreting our results. First, our test analyzes only a subset of individuals --- those whose race is differently perceived across multiple encounters. Tests for bias often derive their estimates from a subset of individuals --- for example, the ``veil of darkness'' test analyzes individuals stopped near sunset~\cite{grogger2006}, and outcome tests~\cite{simoiu2017,ayres2002outcome} analyze only individuals searched by the police. Further, the population we analyze (people whose race is inconsistently perceived) themselves constitute a population of substantial interest~\cite{harris2002, maclin2001, herman2010, shih2009, gaither2015}. Second, while under our model the officer first perceives an individual's race, and \emph{then} decides whether to search them, these two judgments could also occur in the opposite order. This reverse order seems less likely, given previous work showing that racial judgments occur extremely quickly~\cite{ito2003race,amodio2014neuroscience}. However, this reverse order would add nuance to the interpretation of our results, since it would suggest that the officer's decision to search or arrest the driver caused them to perceive the driver as Hispanic, rather than the other way around. Importantly, this would also be a concerning bias --- namely, that the officer's perception of Hispanic people includes ``people I search and arrest'', which likely leads them to engage in statistical discrimination when deciding whether to search or arrest. We also note that such ambiguities are common in tests for bias: for example, outcome tests can show bias either because officers mis-estimate the probability that drivers carry contraband, or correctly estimate the probability and apply different thresholds when searching drivers of different races. 

A third important point to bear in mind when applying our method is that it requires data on \emph{perceived race} as opposed to \emph{self-identified race}. We hence verify, by reaching out to the state patrols of all three states whose data we analyze, that race data is recorded according to the perception of the officer. In other settings, we similarly recommend reaching out to domain experts familiar with the data recording process prior to applying our method. Finally, past work shows that police may sometimes strategically misclassify minorities as white to conceal disparate rates of negative treatment~\cite{pierson2020,luh2022not,friberg2015}. This misclassification would lead us to \emph{underestimate} the extent of bias against Hispanic drivers (since officers will be more likely to strategically misclassify them as white when searching them). Hence, even in the presence of strategic misclassification, our finding of bias against Hispanic drivers likely holds. (We also minimize the impact of this misclassification by filtering out time periods where past work suggests it occurred; see SI for full details).

While we apply our method to assess racial bias in policing, our approach is broadly applicable to testing for bias in datasets which 1) track the same individual over time and 2) record perceptions of the person’s race or other sensitive attributes, like gender. Such datasets occur in many other settings~\cite{saperstein2016making,roth2016multiple}, including healthcare~\cite{samalik2023discrepancies,mcalpine2007agreement,hasnain2010barriers}, surveys~\cite{saperstein2016making,saperstein2006double}, criminal justice~\cite{finlay2024implications}, and child welfare~\cite{baron2024unwarranted}, representing a rich set of directions for future work.

\bmhead{Acknowledgements} The authors thank Serina Chang, Sam Corbett-Davies, Natalia Emanuel, Nikhil Garg, Allison Koenecke, Shengwu Li, Rajiv Movva, and Evan Rose for helpful conversations which informed this manuscript; Johann D. Gaebler, Sharad Goel, Cheryl Phillips, and Vignesh Ramachandran for assistance with the Open Policing data; and the Arizona Department of Public Safety, Texas Department of Public Safety, and Colorado State Patrol for guidance about state policing practices. This work was supported by a Google Research Scholar award, NSF CAREER \#2142419, a CIFAR Azrieli Global scholarship, a LinkedIn Research Award, and the Abby Joseph Cohen Faculty Fund.

\backmatter

\bibliography{sn-bibliography}

\begin{appendices}

\section{Data processing}\label{sec:si_data_processing}

Our analysis relies on data and data processing procedures from the Open Policing Project~\cite{pierson2020}. We analyze state patrol stops from Arizona (2010-2015), Colorado (2010-2016), and Texas (2016), which are the three states which provide the requisite data to match the same driver across multiple stops. Analysis was deemed exempt from review by the Cornell IRB (IRB0148948).

\subsubsection*{Processing of driver race} We follow procedures used in~\cite{pierson2020} to standardize the driver race categories used in the original raw policing data. Our only deviation from the original authors' race processing procedures occurs in Texas. In late 2015, investigative journalists provided evidence that Texas police were misclassifying Hispanic drivers as white~\cite{friberg2015}, something the agency subsequently corrected. To correct for this issue, the original Open Policing authors imputed Hispanic ethnicity from driver surname, and re-categorized drivers who appeared to have been incorrectly classified by police. However, since we wish to analyze the officer's \emph{perception} of driver's race, we do not perform this re-categorization, since this might obscure instances where the officer truly perceived the driver as white. We also remove data recorded prior to 2016, since the evidence suggests that data contains instances of deliberate misrecording (as opposed to good-faith misperception) by the police. From 2016 onwards, the data recording by the police appears much more reliable: specifically, the fraction of drivers with Hispanic surnames who are recorded as white falls sharply in December 2015, just after the investigative journalists published their report. This gives us confidence that data from 2016 onward is more reliably recorded, and in particular that inconsistencies in the recording of driver race result from good-faith misperception as opposed to deliberate misrecording.

Our method requires data on \emph{perceived race} as opposed to \emph{self-identified race}. We hence verified, by reaching out to the relevant departments of all three states whose data we analyzed, that race data is recorded according to the perception of the officer. The Arizona Department of Public Safety confirmed that the data represented officer perception: ``we rely on the trooper's perception, utilizing their best judgement under the circumstances." The Colorado State Patrol told us that ``[t]he race and ethnicity data is in fact the trooper's perception." The Texas Highway Patrol provided us with an excerpt from their policy manual which instructs officers to ``[u]se your best judgement in determining the race or ethnicity of the individual".\footnote{We note that the situation in Texas is somewhat more ambiguous, because in November 2015, several news articles~\cite{friberg2015, baumgartner2015} suggested that Texas officers were instructed to ask for a driver's race. However, the response we received from the Texas Highway Patrol, combined with the fact that driver race was fairly frequently inconsistently recorded in our data even over a short time period, suggests that this was not in fact consistently done, and that race was recorded based on officer perception. We confirm that our results are robust to excluding data from Texas.} Overall, our investigations confirm that the recorded data represents officer perception in all three states.

\subsubsection*{Matching drivers across multiple stops} 

The data available to match the same driver across multiple stops varies by state. In Arizona, we use the first and last name of the driver and their vehicle style and year, which are available for 99\% of stops; in Colorado, the driver's first and last name and date of birth, which are available for 85\% of stops; and in Texas, the driver's first and last name and complete address, which are available for 96\% of stops.

Our analysis requires the fields we use to correctly differentiate distinct drivers (i.e., two different people should very rarely have the same values). In Texas, this is very likely to be true: two different people are very unlikely to have the same full name and full address. In Colorado, past analyses of the probability two people share the same full name and birth date~\cite{goel2020one} also implies this is very likely to be true in a sample of our size. 
 In Arizona, we are not aware of past work analyzing the probability two distinct people share the same first name, last name, vehicle style, and vehicle year. However, the descriptive statistics of our dataset suggest that these fields are indeed sufficient to differentiate distinct drivers. In particular, if distinct drivers were frequently being incorrectly combined in Arizona, we would expect its fraction of drivers with multiple stops to be higher, and its number of stops for each driver to be higher, than in Colorado, where we have access to fields which reliably differentiate distinct drivers and whose data spans a similar time period. In fact, we observe the opposite: only 15\% of drivers are stopped multiple times in Arizona, as opposed to 22\% in Colorado, and the number of stops per driver in Arizona is 1.2, as opposed to 1.4 in Colorado. This suggests that incorrectly matched drivers are not skewing Arizona data. As a final precaution against incorrectly matched drivers, we remove the small proportion (0.1\%) of drivers who are recorded as having more than 10 stops, to prevent drivers who have been incorrectly matched many times from skewing the results. This does not affect our conclusions.

\pagebreak

\section{Supplementary Figures}

\setcounter{figure}{0}
\renewcommand{\figurename}{Fig.}
\renewcommand{\thefigure}{S\arabic{figure}}

\FloatBarrier

\begin{figure}
    \centering
    \includegraphics[width=0.75\linewidth]{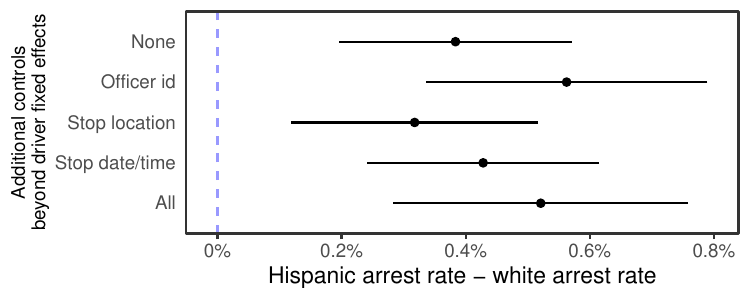}
    \caption{Estimated increase in the \emph{arrest} rate (as opposed to the search rate, as in our primary specification) when the same driver is perceived as Hispanic as opposed to white, using a linear probability model. All estimates include driver fixed effects. 95\% confidence intervals are plotted with standard errors clustered at the driver level. Estimates use Colorado and Arizona data because Texas does not provide arrest data. The finding of bias against Hispanic drivers remains robust when using this alternate outcome, and including controls for officer, stop location, and stop date/time.}
    \label{fig:az_co_arrest_rate_fig}
\end{figure}

\begin{figure}
    \centering
    \includegraphics[width=0.75\linewidth]{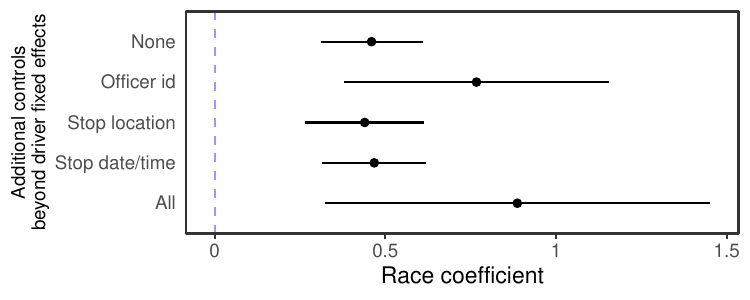}
    \caption{Estimates from a fixed effects generalized linear model with a logit link. The horizontal axis plots the coefficient on driver race = Hispanic after controlling for driver fixed effects. The finding of bias against Hispanic drivers remains robust when using this alternate statistical model, and including controls for officer, stop location, and stop date/time.}
\label{fig:feglm_no_bias_correct_search_rate_fig}
\end{figure}

\begin{figure}
    \centering
    \includegraphics[width=0.75\linewidth]{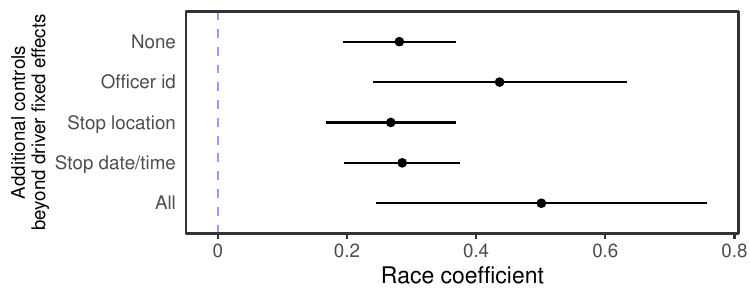}
    \caption{Estimates from a conditional logistic regression model~\cite{breslow1978estimation} with a stratum for each driver. The horizontal axis plots the coefficient on driver race = Hispanic. The finding of bias against Hispanic drivers remains robust when using this alternate statistical model, and including controls for officer, stop location, and stop date/time.}
    \label{fig:cond_logit_search_rate_fig}
\end{figure}

\end{appendices}

\end{document}